\documentstyle[11pt]{article}
\renewcommand{\theequation}{\arabic{section}.\arabic{equation}}
\newcommand{\bm}[1]{\mbox{\boldmath $#1$}}
\begin{document}
\begin{titlepage}
\setcounter{page}{1}
\title{
\hfill {\small GCR--96/07/02}\\
\hfill {\small DTP--MSU/96--11}\\
\bigskip
\bigskip
STATIONARY BPS SOLUTIONS\\ TO DILATON-AXION GRAVITY}
\author{
\bigskip
G\'erard Cl\'ement\thanks{E-mail: gecl@ccr.jussieu.fr.}$\,\,^1$
{\small and}
Dmitri 
%
Gal'tsov\thanks{E-mail: galtsov@grg.phys.msu.su}$\,\,^2$\\
\small $^1\,$ Laboratoire de Gravitation et Cosmologie Relativistes,\\
\small Universit\'e Pierre et Marie Curie, CNRS/URA769 \\
\small Tour 22-12, Bo\^{\i}te 142, 4 place Jussieu,
75252 Paris cedex 05, France \\
\small $^2\,$ Department of Theoretical Physics, Physics Faculty,\\
\small Moscow State University, 119899, Moscow, Russia
}
\bigskip
\date{July 2, 1996}
\maketitle

\begin{abstract}
Stationary four-dimensional BPS solutions to gravity coupled bosonic
theories admitting a three--dimensional sigma--model representation on
coset spaces are interpreted as null geodesics of the target manifold equipped
with a certain number of harmonic maps. For asymptotically flat (or Taub--NUT)
space--times such geodesics can be directly parametrized in terms of
charges saturating the Bogomol'nyi--Gibbons--Hull bound, and classified
according to the structure of related coset matrices. We investigate in
detail the ``dilaton--axion gravity'' with one vector field,
and show that in the space of BPS solutions an $SO(1,2) \times SO(2)$
classical
symmetry is acting. Within the present formalism the most general
multicenter (IWP/Taub--NUT dyon) solutions are derived
in a simple way. We also discover a large new class of asymptotically flat
solutions for which the dilaton and axion charges are constrained only
by the BPS bound. The string metrics for these solutions are generically
regular. Both the IWP class and the new class contain massless solutions.
\vskip5mm
\noindent
PACS number(s): 97.60.Lf, 04.60.+n, 11.17.+y

\end{abstract}

\end{titlepage}

\section{Introduction}

A typical feature of many non--linear field theories is the existence
of certain positivity bounds for the energy similar to the
Bogomol'nyi--Prasad--Sommerfield ( BPS) bound \cite{b} in the
Yang--Mills--Higgs models admitting monopoles. Classical solutions
saturating these bounds (BPS solutions) possess topological
charges and minimize the energy for
given values of these charges. Consequently they are stable,
and usually associated with solitons. Physically
the BPS saturation condition means that forces between them are balanced,
so generically the corresponding
multisoliton solutions also exist.
In supersymmetric embeddings solitons possess unbroken
supersymmetries \cite{wo} which are manifest in the
existence of
Killing 
spinors constant at spatial infinity. The vanishing of the corresponding 
spinorial variations on soliton backgrounds gives rise to linear
Bogomol'nyi type equations which facilitate the effective construction of
these solitons. Topological charges enter
as central charges into the supersymmetry algebra and hence the knowledge
of the soliton spectrum is important for a proper understanding of
the dynamics of the theory.

Similar phenomena are also well known in theories including gravity.
The Einstein theory can be embedded in simple supergravity, the
corresponding positivity argument providing a 
concise 
proof of the positive mass conjecture \cite{wi}. The importance
of BPS bounds in extended supergravities was advocated by
Gibbons \cite{gi2}. In $N=2$ supergravity (containing in the bosonic sector
the graviton and an Abelian vector field) the extreme
Reissner--Nordstr\"om black hole saturates the BPS bound and
has an unbroken supersymmetry, while the general non--singular
solution of the corresponding spinorial equation is the
Majumdar--Papapetrou multicenter solution to the Einstein--Maxwell
(EM) equations \cite{gh}. Since quantum corrections
to supersymmetric backgrounds are controllable, the BPS solutions
are an important tool to investigate the underlying quantum
theory non--perturbatively \cite{rk2}.

The $N=2$ theory, however, has anomalies, so the non--renormalization
argument is not fully applicable. The situation is better in
$N=4$ four--dimensional supergravity, containing in the bosonic
sector the graviton, two scalar fields (dilaton and axion) and six
$U(1)$ vectors. The BPS bound for static asymptotically flat
solutions to this theory was discussed by Gibbons \cite{gi4}.
Truncated models of this kind containing one
(dilaton--axion gravity), or two vector fields were extensively
studied recently by Kallosh and collaborators \cite{rk4} \cite{bko}.
In \cite{kkot} some stationary BPS solutions to dilaton--axion
gravity were given and the BPS condition was generalized to
include the NUT charge. Recently such solutions have been investigated in
the context of heterotic string theory, which gives rise in four
dimensions to a more general bosonic lagrangian including 28 vector
fields, the dilaton and the axion and 132 scalar moduli fields
forming a coset $SO(6, 22)/S(O(6)\times O(22))$. The BPS bound
in this theory was discussed in \cite{ssd}, some static
spherically symmetric solutions saturating this bound were
studied in \cite{cv}. Of particular interest are
massless BPS states, which can substantially affect the low--energy
dynamics \cite{msls}.

In all listed theories there are internal symmetries relating
configurations depending effectively on three space--time
coordinates ({\em e.g.} static or stationary), which can be
used to generate new explicit solutions. This was
one of the main technical tools in most of the above papers.
Another widely used approach was dimensional reduction
>from ten--dimensional solutions known to be exact string
backgrounds, such as stringy plane waves \cite{kkot}. Both methods
generically give physically interesting, but nevertheless
restricted classes of solutions. A more general approach
consists in the direct use of the first order
equations for the supercovariantly constant spinors. Along these lines
Tod was able to find all stationary BPS solutions
to EM theory, dilaton gravity, and a large (although restricted)
class of solutions to $N=4$ supergravity \cite{tod}.
However, the complete enumeration of all stationary asymptotically
flat solutions to dilaton--axion gravity (and larger theories
including moduli fields) seems not to have been given so far.
Here we make some progress in this direction using an alternative
(purely bosonic) way to classify the BPS solutions.

The method utilizes the $\sigma$--model structure of the
three--dimensional theories describing four--dimensional
backgrounds admitting a non--null Killing vector. For pure
gravity this is the well--known Ernst $SL(2, R)/SO(2)$
$\sigma$--model \cite{er}, generalized to Maison's $SL(n-2, R)/SO(n-2)$
$\sigma$--model \cite{ms} for $n$--dimensional vacuum
Kaluza--Klein (KK) theory, while for the EM theory this is
the Neugebauer and Kramer $SU(2, 1)/S(U(2)\times U(1))$
model \cite{nk}, \cite{mg}. More general classes of such
models following from supergravities were studied by
Breitenlohner, Maison and Gibbons \cite{bgm}. The particular
case of dilaton--axion gravity (with one vector field)
was studied in \cite{g}, where the  coset structure
$Sp(4, R)/U(2)$ of the corresponding target space was
found. Typically, when the Killing vector, with respect
to which a reduction to three dimensions is performed,
is timelike, the resulting target space has an indefinite
metric and contains null geodesic lines.
\footnote{Note that the canonical
procedure gives rise in the case of dilaton-axion gravity to the coset 
$Sp(4,R)/U(1,1)$ \cite{bgm}, 
but the indefinite metric can also be presented in terms of
symmetric matrices belonging to $Sp(4,R)/U(2)$ which are somewhat
simpler. In what follows we assume this representation of the non--compact
target space of stationary dilaton--axion gravity.}
It is well known
that geodesics of the target space equipped with some
harmonic functions on a three--space generate a solution
to the $\sigma$--model equations \cite{nk}, \cite{ex}. It was observed in
\cite{gc} that {\em null} geodesics of the target space of stationary
five-dimensional KK theory may be used to generate multisoliton
solutions similar to the Israel-Wilson-Perj\`es solutions \cite{IWP} of EM
theory. The association of stationary multisoliton systems with null
geodesics in target space is a general feature of $\sigma$--models, due to
the fact that in the null case the three--space is conformally flat.
This property can be related to the BPS force balance condition.
Therefore, the classification of stationary BPS solutions
may be traced to the investigation of the null geodesic structure
of the three--dimensional $\sigma$--model target space.

We discuss the general features of this approach in Sect. 2, and then
consider in detail the stationary dilaton--axion gravity model in Sect. 3.
It is shown that isotropic geodesics in the space
$Sp(4, R)/U(2)$ fall into two classes depending on whether the
the corresponding generating matrix is degenerate or not.
The space of classical BPS solutions exhibits a symmetry
under $SO(1, 2)\times SO(2)$ transformations mapping one asymptotically
Taub--NUT geodesic solution onto another
(Sect. 4). These transformations include electric--magnetic duality,
mass -- NUT duality as well as two special boost transformations
in the non--degenerate case. Solutions related to degenerate
matrices are shown to reproduce all previously known BPS solutions including
rotating dilaton--axion Taub--NUT dyon \cite{gk} and dilaton--axion
IWP solutions \cite{kkot}, generalized to
include all possible charges explicitly and to allow for arbitrary
directions of rotation axes. Although the Einstein metrics for these
extreme black hole solutions are singular, the corresponding string
metrics are generically regular in the non-rotating case. Massless states
with non-zero NUT-s are found.
Non--degenerate matrices correspond to formally quasi--periodic solutions
(Sect. 6). These solutions have dilaton
and axion charges constrained only by the BPS condition. Their Einstein
metrics present weak naked singularities, while the associated string
metrics are again generically regular.
Here we present these entirely new solutions in some detail, in view of
their amazing
simplicity. In Appendix A we describe a useful decomposition of the
symplectic algebra, and find some intrinsic connection with 1+2
Clifford algebras. In Appendix B the IWP solutions to the EM
theory are rederived within the same approach.

\section{General formalism}
\setcounter{equation}{0}
 Consider a four--dimensional action
\begin{equation}
S=\int\left( -\frac{R}{16\pi}+L(A^I,\phi^a)\right)\sqrt{-g}d^4x,
\end{equation}
where $A^I$ denotes the set of $U(1)$ (linear) vector fields, and
$\phi^a$ is the set of scalar fields typically including
the dilaton, Peccei--Quinn
axion and moduli. In extended supergravities/superstring effective
theories scalar fields usually form coset spaces, while vector
fields transform under a representation of the corresponding
global symmetry group. When the theory is reduced to three dimensions,
vectors can be traded for pairs of scalar fields, which can sometimes
be incorporated into a larger coset space. The corresponding
three--dimensional theory then turns out to be a gravity coupled
sigma--model on a symmetric space. A general discussion and a list of
different group combinations  ensuring such
a property were given by Julia \cite{ju} and
Breitenlohner, Maison and Gibbons \cite{bgm} and discussed later
in a number of papers.
Assuming a certain familiarity with this formalism, we just briefly
recall here the basic formulas with an emphasis on
the so--called geodesic solutions.

We will be interested in stationary solutions depending
on three (space--like) coordinates, or, in other words, admitting
an (asymptotically) timelike Killing vector field.
Then the standard dimensional reduction ansatz
\begin{equation} \label{dimred}
g_{\mu\nu} = \pmatrix{
f & -f\omega_i \cr
-f\omega_i & -f^{-1}h_{ij}+f\omega_i \omega_j}
\end{equation}
leads to the corresponding three--dimensional theory.
In the stationary case, $U(1)$ vector fields may be expressed in terms
of electric $v^I$ and magnetic $u^I$ potentials, while the rotation
one-form $\omega=\omega_idx^i$ can be dualized to give the twist
potential $\chi$, the pair $(f, \chi)$ forming a coset space
$SL(2, R)/SO(2)$ \cite{nk}. To make $u^I$ and $\chi$
dynamical variables one has to introduce Lagrange multipliers
ensuring the Bianchi identities. The resulting action will be that
of the gravity coupled three--dimensional $\sigma$--model
\begin{equation}
S_\sigma=\frac{1}{2}\int \left({\cal R}-
G_{AB}\partial_iX^A\partial_j
X^B h^{ij}\right)\sqrt{h}\,d^3x,
\end{equation}
where ${\cal R}$ is the Ricci scalar built out of $h_{ij}$,
and the set of scalars $X^A=(f, \chi, v^I, u^I, \phi^a)$, combined into
the potential space endowed with the metric $G_{AB}(X)$, acts as a source
of three--dimensional Euclidean gravity. The equations of motion
for the $X^A$
\begin{equation} \label{hm}
\frac{1}{\sqrt{h}}\partial_i\left(\sqrt{h}h^{ij}G_{AB}
\partial_j X^B\right)=0
\end{equation}
define a harmonic map from the three--space $\{x^i\}$ to the potential
space $\{X^A\}$ (target space).

Obviously, the equations of motion (\ref{hm}) are preserved by
the isometry group $G$ of the target manifold. Therefore this
group is a symmetry group of the space of solutions, using which
new solutions may be generated. In the class of theories under
consideration this group is large enough to ensure the symmetric
space property of the target space, in which case the latter
may be regarded as a coset space $G/H$, with $H$ a subgroup of $G$.
This implies the existence of a matrix $M$ (an appropriate
coset representative) in terms of which the target space metric is
given by
\begin{equation} \label{dl}
dl^2=G_{AB}\,dX^AdX^B=-\frac{1}{4}{\rm Tr}\left(dMdM^{-1}\right).
\end{equation}
Consequently, the three--dimensional action will read
\begin{equation}
S_\sigma=\frac{1}{2}\int \left\{{\cal R}+ \frac{1}{4}
{\rm Tr}\left( \bm\nabla M \bm\nabla M^{-1}\right)\right\}\sqrt{h}\,d^3x,
\end{equation}
where $\bm\nabla$ stands for the three--dimensional covariant derivative,
and a scalar product with respect to the metric $h_{ij}$ is understood.
The corresponding equations of motion take the form of the conservation
equation for the matrix current
\begin{equation} \label{mc}
\bm\nabla\left(M^{-1}\bm\nabla M\right)=0,
\end{equation}
while the three--dimensional Einstein equations read
\begin{equation} \label{ei}
{\cal R}_{ij}=-\frac{1}{4}{\rm Tr}\left(\nabla_i M \nabla_j M^{-1}\right).
\end{equation}

As was noticed by Neugebauer and Kramer \cite{nk}, when one makes the
special assumption that all target space coordinates $X^A$ depend on
$x^i$ through only one scalar potential, i.e. $X^A=X^A[\sigma(x^i)]$,
it follows from the equation of motion (\ref{hm}) that this potential
can be chosen to be harmonic,
\begin{equation}
\Delta\sigma=0, \qquad \Delta={\bm\nabla}^2,
\end{equation}
Eq. (\ref{hm}) reducing then to the geodesic equation on the
target space
\begin{equation}
\frac{d^2X^A}{d\sigma^2}+\Gamma^A_{BC}\frac{dX^B}{d\sigma}
\frac{dX^C}{d\sigma}=0.
\end{equation}
Rewriting this in matrix terms,
\begin{equation}
\frac{d}{d\sigma}\left(M^{-1}\frac{dM}{d\sigma}\right)=0,
\end{equation}
one may express the solution to the geodesic equation
in the exponential form
\begin{equation} \label{AB}
M = A\,{\rm e}^{B\sigma},
\end{equation}
where $A \in G/H$ and $B\in {\it {\cal L}ie}(G)$ are constant matrices
such that $M\in G/H$. Usually we are interested in asymptotically flat
(Taub--NUT) solutions, and it is assumed that $\sigma(\infty)=0$.
Then the asymptotic value of $M$ is $A$, some fixed
quantity depending on the choice of the coset representative.
Asymptotically flat solutions of this kind are thus
target space geodesics passing through the point $A$.

The three--dimensional Einstein equations (\ref{ei})
read, in the case of geodesic solutions,
\begin{equation}
{\cal R}_{ij}=\frac{1}{4}{\rm Tr}(B^2)\nabla_i \sigma \nabla_j \sigma.
\end{equation}
\noindent From this expression it is clear that in the particular case
\begin{equation} \label{null}
{\rm Tr}(B^2)=0
\end{equation}
the three--space is Ricci--flat. In three dimensions the Riemann
tensor is then also zero, and consequently the three--space is flat. From
the Eq. (\ref{dl}) one can see that the condition
(\ref{null}) corresponds to
null geodesics \cite{gc} of the target space
\begin{equation}
dl^2=\frac{1}{4}{\rm Tr}(B^2)\,d\sigma^2=0.
\end{equation}
For different particular theories it was observed that null geodesics
describe extremal black holes satisfying a Bogomol'nyi
type condition on the relevant charges.
The flat nature of the three space supports the anticipation that
the attractive/repulsive forces associated with different charges
are mutually balanced. In cases when the bosonic action
can be embedded into some supergravity, it was argued that this condition
corresponds to the existence of Killing spinors
\cite{rk2}\cite{tod} ensuring the
unbroken supersymmetries of bosonic solutions (usually called
BPS solutions). Although no general proof is known so far
that any BPS solution can be presented as a null geodesic
passing through $A$, at least the inverse statement
seems to be rather plausible \cite{bgm}.

Whereas the group $G$ exhibits a symmetry of the full solution space,
only a restricted set of $G$--transformations
preserves the class of asymptotically flat geodesic solutions.
To find the corresponding subgroup (in particular, we are interested
in symmetries of the BPS subspace), consider a similarity
transformation for the matrix $B$
\begin{equation} \label{BU}
B\rightarrow B_U=UBU^{-1},\quad U\in G,
\end{equation}
and construct the corresponding
\begin{equation}
M_U = A\,{\rm e}^{B_U\sigma}.
\end{equation}
Since $A$ is not transformed, $M_U$ is not necessarily
an element of $G/H$. However for some $U$ it may occur that $M_U\in G/H$
indeed, and thus {\it is} a solution. Such $U$ will be the symmetry
transformation mapping one geodesic solution onto another.
The condition ${\rm Tr}(B_U^2) = 0$ will be satisfied automatically, if
(\ref{null}) holds. Therefore we deal with symmetries of the BPS space
as well.

The various null geodesic solutions fall into equivalence
classes, corresponding to the inequivalent types of matrices $B$
satisfying (\ref{null}). There is only one matrix type in the EM case,
leading to the IWP class (see Appendix B). Three distinct matrix types
are associated with null geodesics of the five-dimensional KK theory
\cite{gc}. The first type yields regular quasi-periodic solutions, the
regular solutions of the second type are extreme black holes, while the
third class contains the KK magnetic monopoles and multipoles \cite{GPS}.
As we shall see in the next section, only two matrix types occur  in
dilaton-axion gravity with one vector field.

The construction (\ref{AB}) may be generalized \cite{gc} to the case of
several harmonic functions $\sigma_a$,
\begin{equation}
\Delta \sigma_a =0 \,,
\end{equation}
by showing that
\begin{equation} \label{mupt}
M = A \exp (\sum_a B_a \sigma_a)
\end{equation}
solves the field equations (\ref{mc}) provided that the commutators $[B_a,
B_b]$ commute with the $B_c$:
\begin{equation} \label{com}
[\,[B_a, B_b], B_c] = 0 \,.
\end{equation}
To prove this, we first rewrite (\ref{mupt}) as
\begin{equation}
M = A\,\exp \,(
- \frac{1}{2} \sum_{c>b} \sum_b [B_b, B_c] \,\sigma_b \sigma_c
)\, \prod_a {\rm e}^{B_a \sigma_a}
\end{equation}
where the first exponential may be treated as a c-number function. We then
obtain for the matrix current
\begin{equation} \label{current}
M^{-1} \nabla M = \sum_a B_a \nabla \sigma_a - \frac{1}{2}
\sum_{c>b} \sum_b [B_b, B_c] (\sigma_b \nabla \sigma_c - \sigma_c \nabla
\sigma_b) \,,
\end{equation}
which is conserved if the $\sigma_a$ are harmonic. The three-dimensional
Einstein equations (\ref{ei}) generalize to
\begin{equation}
R_{ij} = \frac{1}{4} \,\sum_a \sum_b {\rm Tr}(B_a B_b)
\,\nabla_i\sigma_a \nabla_j\sigma_b \,,
\end{equation}
so that the three-space is Ricci flat if the matrices $B_a$ satisfy
\begin{equation} \label{bal}
{\rm Tr}(B_a B_b) = 0 \,.
\end{equation}
So the number of independent harmonic functions on which may depend a BPS
solution of the form (\ref{mupt}) is limited by the number of independent
mutually orthogonal null vectors of the target space. In the class
of theories we are dealing with, the target space is a locally Minkowskian
space $M_{p, q}$, where $p$ counts the positive eigenvalues coming from
the two gravitational potentials $f$, $\chi$ and the effective scalar
fields, and $q$ is the number of negative eigenvalues
coming from electric and magnetic potentials. Then the number of
independent null vectors is $\inf (p, q)$. The actual number of possible
independent potentials, however, may be less because of the extra
condition  (\ref{com}).

\section{Dilaton-axion gravity}
\setcounter{equation}{0}
In the following we turn to a specific theory which may be regarded
as a (truncated) bosonic sector of the four--dimensional
heterotic string effective action, or that of the $N=4, \, D=4$
supergravity with only one non--zero vector field. The model describes
the gravity--coupled system of two scalar fields (dilaton $\phi$ and
(pseudoscalar) axion $\kappa$), and one Abelian vector field $A_{\mu}$:
\begin{equation} \label{an}
S=\frac{1}{16\pi}\int \left\{-R+2\partial_\mu\phi\partial^\mu\phi +
\frac{1}{2} e^{4\phi}
{\partial_\mu}\kappa\partial^\mu\kappa
-e^{-2\phi}F_{\mu\nu}F^{\mu\nu}-\kappa F_{\mu\nu}{\tilde F}^{\mu\nu}\right\}
\sqrt{-g}\,d^4x,
\end{equation}
where ${\tilde F}^{\mu\nu}=\frac{1}{2}E^{\mu\nu\lambda\tau}F_{\lambda\tau},\;
F=dA\;$.

Assuming stationarity, one can perform reduction to three
dimensions using the metric ansatz (\ref{dimred}). The vector field
can be parametrized by 2 real functions: an electric
potential $v$,
\begin{equation}
F_{i0}=\partial_iv/\sqrt{2},
\end{equation}
and a magnetic potential $u$,
\begin{equation}
e^{-2\phi}F^{ij}+\kappa {\tilde F}^{ij}=f\epsilon^{ijk}
\partial_ku/\sqrt{2h}.
\end{equation}
The corresponding spatial vector potential $A_i$
may be recovered from $u$ and $v$ by
\begin{equation}
\sqrt{2}\,dA=e^{2\phi}f^{-1}\ast (du-\kappa dv) -dv\wedge\omega.
\end{equation}

To satisfy the mixed components of the four--dimensional Einstein equations
one has to express $\omega_k$ through the twist potential $\chi$
as follows \cite{gk}
\begin{equation}
\tau_i=\partial_i\chi +v\partial_iu-u\partial_iv,\quad
\tau^i=-f^2\epsilon^{ijk}\partial_j\omega_k/\sqrt{h}.
\end{equation}
Here and below 3--indices are raised and lowered using the three--space
(euclidean signature) metric $h_{ij}$ and its inverse $h^{ij}$.

The resulting three-dimensional gravity coupled $\sigma$--model
has a six--dimensional target space,
$X^A=(f,\,\chi,\,v,\,u,\,\phi,\,\kappa),\;$ $A=1,...,6$ endowed with
the metric
\begin{equation} \label{dl2}
dl^2 =
\frac{1}{2f^2}\left\{ df^2+(d\chi+vdu-udv)^2\right\}
-\frac{1}{f}\left\{e^{-2\phi}dv^2+e^{2\phi}(du-\kappa dv)^2\right\}+
2d\phi^2+\frac{1}{2}e^{4\phi} d\kappa^2 .
\end{equation}

A representation similar to (\ref{dl2}) has been found for the
stationary EM system by Neugebauer
and Kramer \cite{nk}, their formula can be recovered by setting
$\phi=\kappa=0$. As was shown by Mazur \cite{mg}, the EM
target space is isomorphic to the symmetric space
$SU(2,1)/S(U(2)\times U(1))$. When the dilaton and axion fields are
introduced one obtains instead the six--dimensional symmetric space
$Sp(4,R)/U(2)$ \cite{g}. A matrix
representative of this coset can be chosen to be
the symmetric symplectic matrix
 \cite{diak}
\begin{equation} \label{MPQ}
M=\left(\begin{array}{crc}
P^{-1}&P^{-1}Q\\
QP^{-1}&P+QP^{-1}Q\\
\end{array}\right),
\end{equation}
where $P$ and $Q$ are the real symmetric $2 \times 2$ matrices
\begin{equation}
P=-{\rm e}^{-2\phi}
\pmatrix {
v^2-f{\rm e}^{2\phi} & v \cr
v & 1 \cr}, \;\;
P^{-1}= f^{-1}\pmatrix {
1&-v \cr
-v & v^2-f{\rm e}^{2\phi} \cr}, \;\;
Q= \pmatrix {
vw-\chi & w \cr
w & -\kappa \cr},
\end{equation}
with $w=u-\kappa v$.

A solution $M$ depending on only one potential
is a geodesic
\begin{equation} \label{AB3}
M = A {\rm e}^{B\sigma}
\end{equation}
passibg through
a given point $A$. If we choose as usual the harmonic potential to vanish
at spatial infinity, then $A = M(\infty)$. For the asymptotically flat and/or
asymptotically Taub--NUT
configurations we are interested in, the ``potentials'' $X^A$
should be normalized at spatial infinity by
\begin{equation} \label{gauge}
f(\infty)=1,\quad \chi(\infty)=v(\infty)=u(\infty)=\phi(\infty)=
\kappa(\infty)= 0,.
\end{equation}
leading to
\begin{equation} \label{A}
A=\pmatrix{
\sigma_3 & 0  \cr
0 & \sigma_3  \cr}
\end{equation}
(here and below $I$ is the $2\times 2$ unit matrix,
$\sigma_1=\sigma_x,\, \sigma_2=i\sigma_y,\, \sigma_3=\sigma_z,$
and $\sigma_x$, $\sigma_y$, $\sigma_z$ are the standard Pauli matrices
with $\sigma_z$ diagonal).
Now we state the conditions on the matrix $B$ such that $M\in Sp(4,R)/U(2)$.
The symmetry of $M$ leads to the pseudosymmetry condition
\begin{equation} \label{BA}
B^TA = AB.
\end{equation}
$M$ is also symplectic, $M^T JM = J$, with
\begin{equation} \label{J}
J=\pmatrix{
0 & I \cr
-I & 0  \cr}
\end{equation}
On account of (\ref{BA}) and of the symplecticity of $A$,
this leads to the condition
\begin{equation}  \label{BK}
BK + KB = 0,
\end{equation}
with
\begin{equation}
K=\pmatrix{
0 & \sigma_3 \cr
-\sigma_3 & 0 \cr}.
\end{equation}
Finally, $M$ is unimodular, leading to
\begin{equation} \label{TrB}
{\rm Tr}\, B = 0.
\end{equation}

As shown in Appendix A, the exponential ${\rm e}^{B\sigma}$
belongs to the coset $Sp(4, R)/(SO(2)\times SO(1, 2))$, where the $SO(2)$
component is generated by $K$, while the remaining non--compact group is
generated by the $\Sigma_a$,
\begin{equation}
\Sigma_a=\left\{
\pmatrix {
0 & I \cr
-I & 0 \cr}, \quad
\pmatrix {
0 & -\sigma_1 \cr
-\sigma_1 & 0 \cr}, \quad
\pmatrix {
\sigma_1 & 0 \cr
0 & -\sigma_1 \cr}
\right\}
\end{equation}
(note that $\Sigma_0 \equiv J$), with the commutation relations
\begin{equation} \label{Sig}
\left[ \Sigma_a, \;\Sigma_b \right]=2{\varepsilon_{ab}}^c\Sigma_c \,,
\qquad \left[K,\; \Sigma_a \right]=0 \,.
\end{equation}
Here and below we use $SO(1,2)$ indices $a, b, c=0, 1, 2$ which are
lowered/raised using the 1+2 Minkowski metric
$\eta_{ab}={\rm diag}(1, -1, -1)$ and
its inverse $\eta^{ab}=\eta_{ab}$.

The six remaining elements of $sp(4, R)$
can be cast into two $so(1,2)$ (co)vectors
\begin{equation}
{\Gamma^1}_a=\left\{
\pmatrix {
0 & \sigma_1  \cr
-\sigma_1  & 0 \cr}, \quad
\pmatrix {
0 & -I \cr
-I & 0 \cr}, \quad
\pmatrix {
I & 0 \cr
0 & -I \cr}
\right\},
\end{equation}
and
\begin{equation}
{\Gamma^2}_a=\left\{
\pmatrix {
\sigma_2 &  0 \cr
0  & \sigma_2  \cr}, \quad
\pmatrix {
\sigma_3 &  0 \cr
0  & -\sigma_3  \cr}, \quad
\pmatrix {
0 & \sigma_3  \cr
\sigma_3  & 0 \cr}
\right\},
\end{equation}
satisfying the commutation relations (\ref{so2}), (\ref{so12}) with $K$ and
$\Sigma_a$.

Therefore, the most general $B$ can be parametrized by
two $SO(1, 2)$ vectors $\bm\alpha,\, \bm\beta$ as follows
\begin{equation} \label{B}
B=\alpha^a\,{\Gamma^1}_a\;+\;\beta^a\,{\Gamma^2}_a
\equiv \bm\alpha \cdot\bm\Gamma^1+\bm\beta \cdot\bm\Gamma^2.
\end{equation}
To perform exponentiation one needs an anticommutator
\begin{equation}
\left\{{\Gamma^1}_a, \;{\Gamma^2}_b\right\}=-2{\varepsilon_{ab}}^c\;
\eta^{cd}\;{\Gamma^0}_d,
\end{equation}
which lies outside the Lie algebra $sp(4, R)$:
\begin{equation}
{\Gamma^0}_a=\left\{
\pmatrix {
\sigma_3 &  0 \cr
0  & \sigma_3  \cr}, \quad
\pmatrix {
\sigma_2 &  0 \cr
0  & -\sigma_2  \cr}, \quad
\pmatrix {
0 & \sigma_2  \cr
\sigma_2  & 0 \cr}
\right\}.
\end{equation}
Remarkably, this third triplet of matrices makes the full system $SO(1,2)$
covariant on the upper index as well, as discussed in Appendix A.
The anticommutator
\begin{equation} \label{anti}
\left\{{\Gamma^c}_a, \;{\Gamma^d}_b\right\}=
2\left(\eta^{cd}\eta_{ab}\,I - {\varepsilon^{cd}}_f\,{\varepsilon_{ab}}^e\;
{\Gamma^f}_e\;\right)
\end{equation}
is valid for all $c$, $d$. This last relation shows that the three sets
of gamma's form 1+2 Clifford algebras.

Using these formulas one immediately obtains for the square of (\ref{B})
\begin{equation} \label{B2}
B^2=-\left( \bm\alpha^2+\bm\beta^2\right)I-
2\left( \bm\alpha \wedge \bm\beta\right)\cdot {\bm\Gamma}^0,
\end{equation}
so that the null geodesic condition (\ref{null}) reads
\begin{equation} \label{TrB2}
{\rm Tr}(B^2) = -4(\bm{\alpha}^2 + \bm{\beta}^2) = 0.
\end{equation}
Assuming this condition to hold, we obtain successively
\begin{equation} \label{B3}
B^3=-2\left(\bm\alpha'\cdot {\bm\Gamma}^1+\bm\beta'\cdot{\bm\Gamma}^2\right),
\end{equation}
where
\begin{equation}
\bm\alpha'=\bm\beta \wedge(\bm\alpha \wedge \bm\beta), \quad
\bm\beta'= \bm\alpha \wedge (\bm\beta \wedge \bm\alpha),
\end{equation}
and
\begin{equation}  \label{B4}
B^4=4(\bm\alpha \wedge \bm\beta)^2 \,I.
\end{equation}

The fact that the matrix $B^4$ is proportional to unity could also have been
derived from the matrix identity
\begin{equation} \label{PB}
{\cal P}(B) = 0
\end{equation}
where
\begin{equation}
{\cal P}(\lambda)=\det(B-\lambda I)=\prod_{i=1}^4(\lambda-\lambda_i)
\end{equation}
is the characteristic polynomial. Because the matrix $B$ is constrained by
equations (\ref{TrB}) and (\ref{TrB2}), ${\rm Tr}B = 0$ and
${\rm Tr}(B^2) = 0$,
leading to ${\rm Tr}(B^3) = 0$ as well from eq. (\ref{B3}), only zero and
fourth powers of $\lambda$
enter this characteristic polynomial. Consequently, the identity
(\ref{PB}) reads
\begin{equation} \label{det}
B^4+\det B=0.
\end{equation}

Now we are in a position to classify the possible types of matrices $B$
satisfying the null geodesic condition (\ref{TrB2}).
It follows from Eqs. (\ref{B4}), (\ref{TrB2}) and (\ref{det}) that
there are two essentially different types of $B$ matrices:
\begin{enumerate}
\item[i)]
Type 1: degenerate $B,\; \det B=0$.

\noindent This is the case whether 1a: $\bm\alpha$ and $\bm\beta$ are
{\em collinear},
\begin{equation} \label{deg1}
\bm\beta = c\bm\alpha, \quad 0<c<\infty,\qquad {\rm with} \qquad
\bm\alpha^2 = 0
\end{equation}
>from Eq. (\ref{TrB2}),
or 1b: $\bm\alpha = 0$, ${\bm\beta}^2 = 0$ or {\em vice versa}
(Eq. (\ref{deg1}) goes over to this
in the limit $c \rightarrow \infty$ or $c \rightarrow 0$).
For the type 1, $B^2=0$ and hence the expansion of the exponential
$\exp(B\sigma)$ contains only a linear term.

\item[ii)]
Type 2: non--degenerate $B,\; \det B\neq 0$.
\begin{enumerate}
\item[type 2a]: $\bm\alpha$ and $\bm\beta$ are {\em non--collinear},
but both are null, i.e. ${\bm\alpha}^2=0,\;\;{\bm\beta}^2=0$;
\item[type 2b]: neither of $\bm\alpha$ and $\bm\beta$ is null,
that is either
\begin{equation}
{\bm\alpha}^2=-{\bm\beta}^2>0;
\end{equation}
or
\begin{equation}
{\bm\beta}^2=-{\bm\alpha}^2>0.
\end{equation}
\end{enumerate}
\end{enumerate}
As we will see later, there is an internal symmetry of the space
of geodesic solutions mixing subclasses $a$ and $b$ inside
each type, but preserving the distinction between types 1 and 2.

Now we look for BPS solutions depending on {\em two} harmonic
functions $\sigma, \,\tau$
\begin{equation} \label{2pot}
M=A{\rm e} ^{B\sigma + C\tau},
\end{equation}
with $C$ satisfying the same conditions as $B$, and thus
admitting an expansion
\begin{equation}
C=\bm\gamma\cdot\bm\Gamma^1 + \bm\delta\cdot\bm\Gamma^2,
\end{equation}
and, in addition\footnote{The study of the Lie algebra sp(4,R) shows
that the more general condition (\ref{mupt}) on the commutator
$[B, C]$ reduces to $[B, C] = 0$.},
\begin{equation}
[B,\,C]=0,\quad {\rm Tr}(BC)=0.
\end{equation}
These last equations impose the following conditions on the
$SO(1, 2)$ vectors $\bm\gamma,\, \bm\delta$:
\begin{eqnarray} \label{comp}
\bm\alpha \wedge \bm\gamma + \bm\beta \wedge \bm\delta &=& 0,\nonumber\\
\bm\alpha \cdot \bm\delta - \bm\beta \cdot \bm\gamma &=& 0,\\
\bm\alpha \cdot \bm\gamma + \bm\beta \cdot \bm\delta &=& 0\nonumber.
\end{eqnarray}

These conditions lead to very strong restrictions on possible
$\bm\gamma,\, \bm\delta$ once $\bm\alpha,\, \bm\beta$ are given.
Namely, for $B$ of type 1, the matrix $C$ should be also of type 1,
and $\bm\gamma$, $\bm\delta$ should be collinear to $\bm\alpha$,
$\bm\beta$; it then follows that $BC = CB = 0$.
Redefining $\sigma,\, \tau$ through suitable linear
combinations one can then always choose
\begin{eqnarray} \label{BC}
B = (0, \bm\beta), & C = (\bm\beta, 0), & {\rm with}
\;\;\; \bm\beta^2 = 0.
\end{eqnarray}
Since for the type 1 the exponents in (\ref{2pot}) contain only linear terms
it is clear that the most general solution can be presented
in the equivalent form
\begin{equation} \label{lin}
M=A\left\{1+\bm\beta\cdot\left(\bm\Gamma^1\tau+
\bm\Gamma^2\sigma\right) \right\}.
\end{equation}

If $B$ is of type 2, then the only solution is $C\propto B$
(i.e. there is no extreme solution really depending on two
potentials). Since the target space (\ref{dl2}) is locally $M_{4,2}$,
it is also clear that for the present system no BPS solutions
depending on more than two potentials are possible.

\section{Physical charges and $SO(1,2)\times SO(2)$ symmetry of BPS space}
\setcounter{equation}{0}

\noindent From the above analysis it follows that any stationary
asymptotically flat BPS solution
to dilaton--axion gravity can be parametrized by six real numbers
subject to the null--geodesic condition (\ref{TrB2}). These numbers can
be conveniently arranged into two $SO(1, 2)$ vectors
$\bm\alpha,\,\bm\beta$. Here we want to relate them to physical charges
which can be introduced via the standard asymptotic expansions of the
relevant fields in the asymptotically Taub--NUT space--time
(including the usual asymptotically flat one as a particular case).
Here we introduce the charges in accordance with the paper \cite{gk}
as follows (recall the gauge conditions (\ref{gauge}))
\begin{eqnarray} \label{as}
f & \sim & 1-\frac{2M}{r}   ,\nonumber\\
\chi  & \sim &  -\frac{2N}{r}  ,\nonumber\\
v  & \sim &  \frac{\sqrt{2}Q}{r}  ,\nonumber\\
u  & \sim &   \frac{\sqrt{2}P}{r},     \\
\phi  & \sim & \frac{D}{r}   ,\nonumber\\
\kappa  & \sim & \frac{2A}{r}   ,\nonumber
\end{eqnarray}
It is also convenient to use a complex mass combining the Schwarzschild mass
$M$ and the NUT charge $N$:
\begin{equation}
m=M+iN,
\end{equation}
a complex axidilaton charge combining the dilaton $D$ and axion $A$
charges
\begin{equation}
d=D+iA,
\end{equation}
and a complex electromagnetic charge
\begin{equation}
q=Q+iP,
\end{equation}
joining the electric $Q$ and magnetic $P$ charges.

To make the identification of the components of the vectors
$\bm\alpha,\, \bm\beta$ with physical charges, we consider a solution
(\ref{AB}) depending on a single monopole potential $\sigma$, with the matrix
$B$ of the generic form (\ref{B}). We choose the
vectors $\bm\alpha,\, \bm\beta$ to have the dimension of a length (i.e.
the dimension of the charges in (\ref{as})). The dimension of the harmonic
function
$\sigma$ will then be an inverse length, and we choose the normalization
\begin{equation}
\sigma \sim \frac{1}{r}\,.
\end{equation}
Substituting the expansions (\ref{as}) into the matrix $M$ in the form
(\ref{MPQ}),
and comparing with the corresponding linearization of (\ref{AB3}), one
finds the following correspondence
\begin{eqnarray} \label{albe}
\bm\alpha &=& \left(\sqrt{2}P,\;A-N,\;M+D\right)  ,\nonumber\\
\bm\beta &=& \left(-\sqrt{2}Q,\;M-D,\;N+A\right)  .
\end{eqnarray}
The null geodesic (BPS) condition (\ref{TrB2}) $\bm\alpha^2+\bm\beta^2=0$
then assumes its standard form
\begin{equation} \label{BPS}
Q^2+P^2=M^2+N^2+D^2+A^2,
\end{equation}
or, in complex notation,
\begin{equation}
|q|^2=|m|^2+|d|^2 .
\end{equation}

For type 1 and type 2a some additional constraints are imposed
on $\bm\alpha,\, \bm\beta$. Consider first the degenerate type 1.
The collinearity condition (without assuming the BPS constraint (\ref{BPS}))
gives
\begin{equation} \label{coll1}
M^2+N^2=D^2+A^2,
\end{equation}
\begin{equation}
Q(N-A)=P(M-D),
\end{equation}
\begin{equation}
P(N+A) = -Q(M+D)
\end{equation}
(only two of these equations are independent).
Solving these equations with respect to $D,\, A$ one gets
\begin{eqnarray}
D &=& \frac{M\left(P^2-Q^2 \right) -2NQP}{Q^2+P^2},\nonumber\\
A &=& \frac{N\left(Q^2-P^2 \right) -2MQP}{Q^2+P^2},
\end{eqnarray}
or, in complex form
\begin{equation} \label{coll2}
d=-\frac{{\bar m}q}{{\bar q}}.
\end{equation}

Now, if the BPS condition (\ref{BPS}) is imposed, one gets from
(\ref{coll1})
\begin{equation}
Q^2+P^2=2\left(M^2+N^2 \right),
\end{equation}
or ${\bar q}q = 2{\bar m}m$, so that the relation (\ref{coll2}) takes
the ``holomorphic'' form
\begin{equation} \label{hol}
d=-\frac{q^2}{2m}.
\end{equation}
This relation was found for a rotating Taub--NUT dyon in
dilaton--axion gravity in \cite{gk}.
Particular cases of these dyon configurations are the purely
electric type $P=0$ ($\bm\alpha=0$), for which the extremal states
have dilaton and axion charges
\begin{equation}
D=-M, \; A=N,
\end{equation}
and the purely magnetic type $Q=0$ ($\bm\beta=0$), for which one has
\begin{equation}
D=M, \; A=-N.
\end{equation}

Let us now consider the possibility of solutions with zero
Schwarzschild mass. Within the type 1 such non--trivial configurations
may exist if the NUT parameter
is non--zero. In the generic dyon case (type 1a)
the dilaton and axion charges are
\begin{equation} \label{massless1}
D=-\frac{QP}{N}, \; A=\frac{Q^2-P^2}{2N},
\end{equation}
and the BPS condition reads $Q^2 + P^2 = 2N^2$
(magnetic solutions of this kind were given in \cite{udual}).
In non--dyon cases
\begin{equation} \label{massless2}
D=0,\quad A=\pm N,
\end{equation}
where the upper sign coresponds to electric and lower to magnetic solutions.

Now we discuss the relations between charges for solutions
corresponding to non--degenerate $B$. In
the null case 2a one has in addition to the BPS condition (\ref{BPS})
a single constraint
\begin{equation} \label{2a}
MD-NA=\frac{P^2-Q^2}{2}.
\end{equation}
Contrary to type 1, in the case 2a generically
\begin{equation}
M^2+N^2\neq A^2+D^2
\end{equation}
(if the equality holds we come back to the case 1a).
Solving (\ref{BPS}) and (\ref{2a}) we obtain for the dilaton and axion
charges in the case 2a
\begin{eqnarray}
D &=& \frac{M\left(P^2-Q^2 \right) - 2NQP\varrho}{2\left(M^2+N^2\right)},
\nonumber\\
A &=& \frac{N\left(Q^2-P^2 \right) - 2MQP\varrho}{2\left(M^2+N^2\right)},
\end{eqnarray}
where
\begin{equation}
\varrho^2=1-\frac{\left[Q^2+P^2-2(M^2+N^2)\right]^2}{4Q^2P^2}.
\end{equation}
Comparing this with the ``standard'' relations (4.12) one can see their
difference unless $\rho =1$ in which case we come back to the type 1.
Finally, in the non--null non--degenerate type 2b case
the physical charges are constrained only by the BPS
condition (\ref{BPS}).

Now let us discuss the symmetry transformations acting inside the
class of stationary asymptotically Taub--NUT geodesic solutions
to dilaton--axion gravity. As was explained in Sect. 2, we
look for similarity transformations for the matrix $B$ which
preserve the pseudosymmetry condition (\ref{BA}). It is shown in the
Appendix A that if $M\in Sp(4, R)/U(2)$ and $A$ is given by (\ref{A}),
then $AM={\rm e}^{B\sigma}\in Sp(4, R)/(SO(1, 2)\times SO(2))$.
Therefore the subgroup $H'=SO(1, 2)\times SO(2) $ is the desired
symmetry group. In algebraic terms, to preserve the condition (\ref{BA}),
the generator $X\in sp(4, R)$ of the similarity transformation (\ref{BU})
should satisfy
\begin{equation}
X^TA=-AX.
\end{equation}
In view of the decomposition of the $sp(4, R)$ algebra described
in Appendix A, it is clear that $X\in {\cal H'}$ (cf. (\ref{AH})).
In fact this subgroup of the symplectic group has a larger meaning
as that preserving asymptotic flatness (so it is also relevant to
non--null geodesic solutions). It is worth giving a more detailed description
of this symmetry with an emphasis on its action on BPS solutions.

The group $H'$ consists of two components. The first one
($SO(2)$) is generated by $K$, {\em i.e.} in (\ref{BU})
$U={\rm e}^{K\epsilon/2}$. This transformation
mixes $\bm\alpha$ and $\bm\beta$:
\begin{eqnarray} \label{dual}
&&\bm\alpha \rightarrow \bm\alpha \cos\epsilon + \bm\beta \sin\epsilon, \nonumber\\
&&\bm\beta \rightarrow \bm\beta \cos\epsilon - \bm\alpha \sin\epsilon.
\end{eqnarray}
In terms of charges this is an electric--magnetic duality rotation,
{\em i.e.} the transformed quantities are
\begin{equation}
q'=q{\rm e}^{-i\epsilon},\quad  m'=m{\rm e}^{-i\epsilon},\quad
d'=d{\rm e}^{-i\epsilon}.
\end{equation}
It is worth noting that when $Q$ and $P$ get mixed by this transformation,
at the same time the mass and the NUT charge are also mixed, {\em i.e.}
one is not allowed to make a dyon from a purely electric (purely magnetic)
asymptotically flat (NUT-less) configuration without introducing a
NUT charge.
Clearly, the axidilaton charge follows the transformation rule
implied by (\ref{hol}). Comparing this with the general case of
the action of the symplectic group presented as the set of scale,
gauge, $S$-duality, Harrison and Ehlers transformations \cite{diak}, one can
observe that the $K$--transformation is a certain combination of
Ehlers, gravitational gauge and two $S$--duality transformations.

Now consider the $SO(1, 2)$ part of $H'$. It is generated by the
$\Sigma_a$ according to (\ref{Sig}).
In view of (\ref{so12}), these are the transformations
acting independently (and similarly) on $\bm\alpha$ and $\bm\beta$.
Pure `spatial' rotations are given by $U={\rm e}^{\Sigma_0 \theta /2}$:
\[
\alpha^0\rightarrow \alpha^0,
\]
\begin{equation}
\alpha^1\rightarrow \alpha^1 \cos\theta +\alpha^2\sin\theta,
\end{equation}
\[
\alpha^2\rightarrow \alpha^2 \cos\theta -\alpha^1\sin\theta
\]
(similarly for $\bm\beta$).
In terms of charges we find
\begin{eqnarray}
q' & = & q, \nonumber\\
m' & = & m {\rm e}^{-i\theta},\\
d' & = & d {\rm e}^{i\theta}.\nonumber
\end{eqnarray}
This transformation can be used to compensate the mass -- NUT charge
mixing when the $K$--transformation is applied, if strict asymptotic
flatness is desired to be preserved. On the other hand, this
transformation may be used to generate a NUT charge for
purely electric and purely magnetic configurations. It is also
a (different) combination of Ehlers, gravitational gauge and two of
$S$--duality transformations.

The two other $SO(1,2)$ transformations are boost--like, they have different
actions on null and non--null $\bm\alpha$ and $\bm\beta$. Consider
first the non--null case.
The boost in the plane $0-1$ is generated by $U=e^{\Sigma_2 \xi /2}$:
\[
\alpha^0\rightarrow \alpha^0 \cosh\xi-\alpha^1\sinh\xi,
\]
\begin{equation}
\alpha^1\rightarrow \alpha^1 \cosh\xi -\alpha^0\sinh\xi,
\end{equation}
\[
\alpha^2\rightarrow \alpha^2  .
\]
Together with the corresponding transformation of $\bm\beta$ this results in
\begin{eqnarray}
q' & = & q\cosh\xi+\frac{m-d}{\sqrt{2}}\sinh\xi,\nonumber\\
m'+d' & = & m + d,\\
m'-d' & = &(m-d)\cosh\xi + \sqrt{2}q \sinh\xi.\nonumber
\end{eqnarray}
This is a combination of electric gauge and Harrison transformations
which change all six charges in a non--trivial way.
Similarly, the $0-2$ boost $U=e^{\Sigma_1 \eta /2}$ gives
\begin{eqnarray}
q' & = & q\cosh\eta + i\frac{m+d}{\sqrt{2}}\sinh\eta,\nonumber\\
m'- d' & = & m - d,\\
m'+ d' & = &(m+d)\cosh\eta - i\sqrt{2}q \sinh\eta.\nonumber
\end{eqnarray}
This corresponds to a combination of magnetic gauge and Harrison
transformations.

If ${\bm\alpha}$ and ${\bm\beta}$ are null, the boosts reduce
to rotations and rescaling. Hence the action of the full $SO(1,2)$
component of $H'$ is in this case two--parametric. If in some frame
\begin{equation}
\bm\alpha=\alpha^0(1, {\bf n}),
\end{equation}
where ${\bf n}$ is a unit two--dimensional vector,
in any other frame one has
\begin{equation}
\bm\alpha'=\alpha^{'0}(1, {\bf n}'),
\end{equation}
where ${\bf n}'$ is a rotated version of ${\bf n}$. Note that
${\bm\beta}$ should be transformed using the same parameters. Taking also
into account the additional degree of freedom due to the action of $K$, we
see that for a given potential $\sigma$ the general type 1 solution
depends on three parameters (we do not count here for the asymptotic
values of the dilaton and axion fields which can easily be induced
once the solutions in the present gauge are given).

Generally both ${\bm\alpha}^2$ and ${\bm\beta}^2$
as well as the scalar product
\begin{equation}
\bm\alpha\cdot\bm\beta=-2(QP+AM+ND)
\end{equation}
are invariant under the $SO(1, 2)$ transformations.
These quantities, however, are not preserved by the electric--magnetic
duality transformation (\ref{dual}), so if one wishes to identify
solutions related by symmetries, formally there is no distinction between
subcases $a$ and $b$ in the classification of the previous section.
However the distinction still may be useful because the corresponding
solutions are physically different.

In the general case a rescaling
\begin{equation}
\bm\alpha \rightarrow k \bm\alpha,\quad \bm\beta \rightarrow k \bm\beta,
\end{equation}
is also an invariance transformation although trivial, equivalent
to a redefinition of harmonic functions, physically it corresponds
to an equal rescaling of all six charges, {\em i.e.} a change of
length scale.

\section{Type 1 solutions}
\setcounter{equation}{0}
\noindent
All solutions belonging to the degenerate type 1 can be presented
in the form (\ref{lin}), generally they depend on two real potentials
$\sigma$, $\tau$ or one complex potential
\begin{equation}
\zeta=\sigma-i\tau.
\end{equation}
Contrary to the previous section, here we choose this potential to be
dimensionless,
so that the $SO(1, 2)$ null vector $\bm\beta$ in (\ref{lin}) will also be
dimensionless. Normalizing this vector so that $\beta^0=1$,
one can write
\begin{equation} \label{bet}
\bm\beta =(1,\, \cos\alpha,\, \sin\alpha).
\end{equation}
Keeping in mind that from any solution one can generate by $H'$
transformations a two--parameter family of solutions, as
described in the previous section, we could start from a given value of
$\alpha$ (say $\alpha = 0$) in (\ref{bet}). However we find it instructive
to leave here the choice of the gauge $\alpha$ open. The corresponding
solution (\ref{lin}) may be written in complex form
\begin{equation}
M=A\,{\rm Re} \left\{ 1+\bm\beta\cdot
\left(\bm\Gamma^2+i\bm\Gamma^1\right)\zeta\right\}
\end{equation}

Identifying this with the initial coset representation (\ref{MPQ})
we obtain the sigma--model variables in terms of the harmonic potentials
$\sigma$ and $\tau$:
\begin{eqnarray} \label{sol1}
f^{-1} &=& 1+\sigma\cos\alpha+\tau\sin\alpha,\nonumber\\
\chi &=& f(\tau\cos\alpha-\sigma\sin\alpha),\nonumber\\
v+iu &=& -f\zeta,\\
z\equiv\kappa+i{\rm e}^{-2\phi} & = &
i\left(\frac{1+\zeta\cos\alpha}{1+i\zeta\sin\alpha}\right).\nonumber
\end{eqnarray}

Comparing this with the corresponding form of the BPS solutions to the
EM system (see Appendix B) one can observe that
the scale factor $f^{-1}$ now depends linearly on the harmonic functions,
contrasting with a quadratic dependence in the EM case. As we shall see,
this has
a dramatic effect on the regularity property of the solutions.

Let us consider some particular families of BPS solutions
corresponding to different choices of the complex harmonic function
$\zeta$.
\vskip5mm
{\bf Extremal rotating Taub--NUT dyon}
\vskip5mm
We take the harmonic function $\zeta(\bm{r})$ to be
\begin{equation} \label{kerr}
\zeta = - \frac{\sqrt{2}\,q}{|\bm{r} + i\bm{a}|} \,,
\end{equation}
and choose spheroidal coordinates ($r$, $\theta$, $\varphi$) such that
\cite{IWP}
\begin{equation}
(\bm{r} \wedge \bm{a})^2 \equiv a^2(r^2 + a^2)\sin^2\theta \,, \quad
\bm{r} \cdot \bm{a} \equiv ar \cos\theta
\end{equation}
($a = |\bm{a}|$), and $\varphi$ is the azimuthal angle around $\bm{a}$.
Identifying the physical charges by comparing (\ref{albe}) and
(\ref{bet}), and taking into account the relation
\begin{equation} \label{almq}
{\rm e}^{i\alpha} = - \sqrt{2}\,\frac{m}{q} \,,
\end{equation}
which results from (\ref{coll2}), we obtain the solution
\begin{equation} \label{rotm}
ds^2=\frac{\Delta-a^2 \sin^2\theta}{\Sigma}
\left(dt+\omega d\varphi\right)^2
-\Sigma\left(\frac{dr^2}{\Delta}+d\theta^2+
\frac{\Delta\sin^2\theta}{\Delta-a^2\sin^2\theta} d\varphi^2\right),
\end{equation}
where
\[ \Delta=r^2 + a^2 ,\]
\begin{equation}
\Sigma=r^2+a^2\cos^2\theta+2(Na\cos\theta+Mr),
\end{equation}
\[
\omega=\frac{2\left\{(r^2+a^2)N\cos\theta+aMr\sin^2\theta\right\}}
{r^2+a^2\cos^2\theta}.
\]
The electric and magnetic potentials  are
\begin{equation} \label{rotvu}
v=\frac{\sqrt{2}}{\Sigma}\left(Qr+Pa\cos\theta\right),\quad
u=\frac{\sqrt{2}}{\Sigma}\left(Pr-Qa\cos\theta\right),
\end{equation}
and the complex axidilaton is
\begin{equation} \label{rotz}
z=i\left(\frac{r+m-d+ia\cos\theta}{r+m+d+ia\cos\theta}\right).
\end{equation}
The corresponding dilaton factor reads
\begin{equation} \label{rotd}
e^{2\phi}=\frac{1}{\Sigma}
\left|r+2M +ia\cos\theta-\frac{Qq}{m}\right|^2.
\end{equation}

This solution corresponds to the BPS limit of the general rotating
Taub--NUT dyon solution found in \cite{gk} (the radial coordinate used
there differs from the present one by a constant shift $r\rightarrow r-2M$).
For $a\neq 0$ the solution has a naked singularity, as any rotating
BPS solution to this theory in four dimensions \cite{hs}.
For $a=0$ the metric has
a simple form (previously given by Kallosh et al \cite{kkot})
\begin{equation} \label{degmono}
ds^2 = (1+2M/r)^{-1}(\,dt+2N\cos\theta\,d\varphi)^2-
(1+2M/r)\,dr^2 -r(r+2M)(\,d\theta^2+\sin^2\theta\,
d\varphi^2)\,.
\end{equation}
This extreme black hole metric turns out to be singular, the area of the
would-be horizon at $r = 0$ being zero. It is pointed out in \cite{kkot}
that the monopole string metric $d\bar{s}^2 = {\rm e}^{2\phi}\,ds^2$ is
regular for certain parameter values. Actually we have found that it is
always regular, except in the lower-dimensional domain $PM - QN = 0$. The
string metric corresponding to (\ref{degmono}) is
\begin{equation}
d\bar{s}^2 \equiv {\rm e}^{2\phi}ds^2 = \Gamma\left[ \frac{(dt+2N\cos\theta
\,d\varphi)^2}{(r+2M)^2} - \frac{dr^2}{r^2} - d\Omega^2 \right]
\end{equation}
with
\begin{equation} \label{Gamma}
\Gamma = (r+\gamma P)^2 + \gamma^2Q^2, \quad \gamma = 2\frac{PM-QN}{Q^2+P^2}.
\end{equation}
For $Q \neq 0$, $\Gamma$ is positive definite, and the ``throat'' $r = 0$
is at infinite geodesic distance. In the purely magnetic case $Q = 0$,
$\Gamma = (r+2M)^2$ leading
to the same conclusion, except if $M = 0$ (then also $D=0,A=-N,P^2=2N^2$)
where the string metric is massless Taub-NUT. In the
special case $PM - QN = 0$ ($\gamma = 0$) with $M \neq 0$, the string metric
near $r = 0$
\begin{equation}
d\bar{s}^2 \sim r^2(dt + 2N\cos\theta\,d\varphi)^2 - dr^2 - r^2\,d\Omega^2
\end{equation}
has a horizon of zero area at the origin $r = 0$ of Euclidean space. This
``black-point'' singularity \cite{soleng} is mild, all non-radial
geodesics being deflected away from $r = 0$.

The $a \neq 0$ solution depends on the four parameters $\alpha$, $q$
(complex), and $a$. The first three parameters correspond to the three
degrees of freedom of type 1 solutions under the action of the group
$SO(1,2) \times SO(2)$, while the parameter $a$ may be associated with the
scale of the dipole potential $\tau$. However we should keep in mind that
our monopole-dipole solution has been obtained from the choice (\ref{BC})
of the matrices $B$, $C$ in (\ref{2pot}). A more general choice solving
the constraints (\ref{comp}) is, up to $SO(1,2) \times SO(2)$
transformations and rescalings of $\sigma$ and $\tau$,
\begin{equation} B = (0, \bm\beta) \,, \qquad C = (\bm\beta\,\cos\psi,
\bm\beta\,\sin\psi) \,.
\end{equation}
The resulting linear form of the matrix $M$
\begin{equation}
M = A \left\{ 1 + \bm\beta \cdot \left( \bm\Gamma^1 \,\tau\,\cos\psi +
\bm\Gamma^2\,(\sigma + \tau\,\sin\psi) \right) \right\}
\end{equation}
with the same potentials $\sigma$ and $\tau$ as in (\ref{kerr}) leads to
the general monopole-dipole BPS solution depending on five parameters (the
same solution could alternatively be obtained from (\ref{lin}) by
transforming the potentials $\sigma \rightarrow \sigma + \tau\,\sin\psi \,, \tau
\rightarrow \tau\,\cos\psi$).

\vskip5mm
{\bf Multicenter solutions}
\vskip5mm
Let us first discuss the equilibrium conditions between any pair $i,\,j$
of the multicenter system in which each center in endowed with
mass, NUT charge, electric and magnetic charges, and dilaton
and axion charges obeying to (\ref{BPS}), (\ref{coll1}) and (\ref{coll2}).
The condition that the various attractive (tensor and scalar) and
repulsive (vector) forces balance is Eq.(\ref{bal}), which translates into
\begin{equation} \label{bal2}
Q_i Q_j+P_i P_j=M_i M_j+N_i N_j+D_i D_j+A_i A_j
\end{equation}
or, in terms of complex charges,
\begin{equation}
{\rm Re}\left(q_i{\bar q}_j -m_i{\bar m}_j -d_i{\bar d}_j\right)=0.
\end{equation}
Choosing here the general representation of the matrix $B_i$ in terms of
three parameters $\mu_i$, $\alpha_i$, $\psi_i$,
\begin{equation}
\bm\alpha_i = 2\mu_i\,\bm{u}_i\,\cos\psi_i \,\qquad \bm\beta_i =
2\mu_i\,\bm{u}_i\,\sin\psi_i \,,
\end{equation}
where $\bm{u}_i = (1, \cos\alpha_i, \sin\alpha_i)$, Eq.(\ref{bal2}) reads
\begin{equation} \label{bal3}
\bm\alpha_i \cdot \bm\alpha_j + \bm\beta_i \cdot \bm\beta_j \equiv
4\mu_i\mu_j\,\cos(\psi_i - \psi_j)\,\bm{u}_i \cdot \bm{u}_j = 0 \,.
\end{equation}

This condition has two possible solutions. The first solution is $\bm{u}_i
\cdot \bm{u}_j = 0$ which, for null vectors $\bm{u}_i$, $\bm{u}_j$, means
$\alpha_j = \alpha_i$, so that the $\bm\alpha_j$, $\bm\beta_j$ are
collinear to the $\bm\alpha_i$, $\bm\beta_i$; then the whole system of
conditions (\ref{comp}) is satisfied, so that $B_iB_j = B_jB_i = 0$, and
\begin{equation}
M = A\,[1 + B_i\sigma_i + B_j\sigma_j]
\end{equation}
is a two-center BPS solution, which can straightforwardly be
generalized to any number of centers. The condition (\ref{bal3}) is also
solved by $\cos(\psi_i - \psi_j) = 0$, or
\begin{equation} \label{QP}
Q_iQ_j + P_iP_j = 0 \,,
\end{equation}
with $\bm{u}_i \cdot \bm{u}_j \neq 0$; the other conditions (\ref{comp})
are however not satisfied, so that the equilibrium condition (\ref{QP}),
which can hold only for a two-center system, does not allow for a
geodesic BPS solution.

Now we consider an equilibrium configuration of $n$ centers $\bm{r} =
\bm{r}_j$ endowed with complex masses $m_j=M_j+iN_j$ and charges
$q_j=Q_j+iP_j$ (as well as induced axidilaton
charges $d_j=-q_j^2/2m_j$) subject to the condition that the angle
parameter $\alpha_j$ entering the matrix $B_j$, or (by virtue of
(\ref{almq})) the complex ratio $m_j/q_j$, is independent of $j$.
We also
assume that each of these centers carries an arbitrarily oriented dipole
moment $\bm{a}_j$. Then, to write down the multi-rotating Taub-NUT
dyon solution
one merely has to change the complex harmonic function (\ref{kerr}) to
\begin{equation}
\zeta=-\sqrt{2}\sum_{j=1}^n \frac{q_j}{R_j},\qquad
R_j^2=\left({\bm r}-{\bm r}_j+i{\bm a}_j\right)^2 \,.
\end{equation}
This solution generalizes the axion--dilaton IWP solution
presented by Kallosh et al  \cite{kkot} to the dyon case,  presence of
NUT's and arbitrary directions of rotation vectors for each center.
Clearly, for $\bm{a}_j \neq 0$ we deal with naked singularities.

In the non-rotating case, the metric has the form
\begin{equation} \label{multiEMDA}
ds^2=\left( 1+\sum_{j=1}^n \frac{2M_j}{R_j}\right)^{-1}
\left(dt+ \sum_{j=1}^n 2N_j\cos\theta_j\, d\varphi_j\right)^2-
\left( 1+\sum_{j=1}^n \frac{2M_j}{R_j}\right) d\bm{r}^2,
\end{equation}
where $d\bm{r}^2$ is the flat space line element, and $\theta_j$,
$\varphi_j$ are polar angles relative to an arbitrarily oriented
orthogonal frame centered at $\bm{r} = \bm{r_j}$. An essential difference
with the Majumdar--Papapetrou (MP) solution to the EM system
(see Appendix B) is that the metric function $f$ now appears as an inverse
linear function of Coulomb potentials, while the MP
solution (endowed with NUT-s)
\begin{equation} \label{multiEM1}
ds^2= |U|^{-2}(dt + \bm{\omega}\cdot d\bm{r})^2 - |U|^2d\bm{r}^2\,,
\end{equation}
with
\begin{equation} \label{multiEM2}
U = 1 + \sum_{j=1}^n \frac{M_j + iN_j}{R_j}\,, \qquad
\nabla\wedge\bm{\omega} = {\rm Im}\,[2\nabla U + (\bar{U}\nabla U -
U\nabla\bar{U})]\,,
\end{equation}
has quadratic factors. This is the reason why the surfaces $R_j=0$ ,
which for the MP solutions are regular horizons \cite{hah}, correspond
in the dilatonic case to space--time singularities
(the radius of two--spheres shrinks to zero).
It was shown recently \cite{bko} that
adding a second vector field one can get dilatonic solutions sharing
the above MP property too. The other difference between (\ref{multiEMDA})
and (\ref{multiEM1}) is of course the non-linear dependence
(\ref{multiEM2}), in the EM case, of the rotation one-form
$\omega$ on the $N_i$ and $M_i$ (except if all the $N_i$ or all the $M_i$
vanish); this is a consequence of Eq.(\ref{current}) for the non-commuting
matrices $B$ and $C$ of Appendix b.
\vskip5mm
{\bf Massless states}
\vskip5mm
BPS solutions with a vanishing Schwarzschild mass are of particular
interest in the underlying quantum theory since they can substantially
influence the low--energy dynamics \cite{msls}. As was noticed in
the previous section, the type 1 solutions may be massless but at the
expense of a non--zero NUT charge (so that the complex mass $m$
remains finite), the BPS condition now reading $Q^2+P^2=2N^2$.
Although they are not asymptotically flat in the
usual sense, it is worthwhile giving these solutions in explicit form.
All non--rotating massless solutions turn out to possess a locally
flat metric
\begin{equation}
ds^2=(dt+2N\cos\theta\,d\varphi)^2-dr^2-r^2(d\theta^2+\sin^2\theta d\varphi^2),
\end{equation}
which is massless Taub--NUT. The dilaton factor is asymmetric
with respect to magnetic and electric components:
\begin{equation}
{\rm e}^{2\phi}=1-\frac{2QP}{Nr}+\frac{2Q^2}{r^2},
\end{equation}
so for purely magnetic solutions $\phi \equiv 0$, while for purely
electric ones the dilaton field is non--zero, but short--range. In
the general dyon case the dilaton charge is non--zero. The axion field
\begin{equation}
\kappa=\frac{2(Ar-ND)}{(r+D)^2+(N+A)^2}
\end{equation}
is non--zero in all cases, the values of the dilaton and axion charges
being given by (\ref{massless1}) or (\ref{massless2}). In non--dyon cases
the axion field is
short--range.

Massless multicenter solutions may be constructed in a similar way.
In the non--rotating case the metric corresponds
to the massless multi--Taub--NUT space--time
\begin{equation} \label{massless}
ds^2=
\left(dt+ \sum_{j=1}^n 2N_j\cos\theta_j\,d\varphi_j\right)^2-d\bm{r}^2.
\end{equation}
Magnetic solutions of this type
(with zero dilaton field $\phi$) were found earlier \cite{udual}. The metric
(\ref{massless}), as well as the more general metric (\ref{multiEMDA}), may
be asymptotically flat if the sum of the NUT charges
vanishes, $\sum_jN_j = 0$.

\setcounter{equation}{0}
\section{Type 2 solutions}
In the non--degenerate case, the matrix $B$, given by (\ref{B}) where the
non-collinear vectors $\bm\alpha$ and $\bm\beta$ are only constrained by
(\ref{TrB2}), depends on five parameters. One of these parameters is
associated with rescalings of $\sigma$; it is convenient to fix this scale
by the condition
\begin{equation}
\det B =4
\end{equation}
(the determinant of $B$, given by eqs. (\ref{B4}) and (\ref{det}), is
positive by virtue of eq. (\ref{TrB2})).
Then the four non--zero eigenvalues of $B$, solving the characteristic
equation $\lambda_j^4 + 4 = 0$, are
\begin{equation}
\lambda_j=\pm(1\pm i).
\end{equation}
The exponential ${\rm e}^{B\sigma}$ may be computed using the Lagrange
formula
\begin{equation}
{\rm e}^{B\sigma} =\sum_{k=1}^{4}{\rm e}^{\lambda_k \sigma}
\prod_{j\neq k} \frac{B-\lambda_j}{\lambda_k-\lambda_j}.
\end{equation}
Substituting (6.3) into (6.4) one obtains
\begin{equation}
{\rm e}^{B\sigma}=g_0 I+g_+ B + \frac{1}{2}\left(g_1 B^2+g_- B^3\right),
\end{equation}
where
\[ g_0=\cos\sigma\cosh\sigma,\]
\begin{equation}
g_1= \sin\sigma\sinh\sigma,
\end{equation}
\[
g_{\pm}=\frac{1}{2}(\sin\sigma\cosh\sigma\pm \cos\sigma\sinh\sigma).
\]
In the non--degenerate case, BPS solutions to the present system
may depend only on one real potential, so the solution (6.4) is likely
to be the most general one if $\det B\neq 0$.
Using Eqs. (\ref{B}), (\ref{B2}) and (\ref{B3}) one obtains the
exponential as a linear
combination of matrices
${\bm\Gamma}^a=({\Gamma ^a}_0,\,{\Gamma ^a}_1,\,{\Gamma ^a}_1)$
forming a triplet of 1+2 Clifford algebras:
\begin{equation}
{\rm e}^{B\sigma}=g_0 I + \bm\upsilon_a\cdot{\bm\Gamma^a}  ,
\end{equation}
where
\begin{eqnarray}
\bm\upsilon_0 &=& -g_1 \bm\alpha\wedge\bm\beta,\nonumber\\
\bm\upsilon_1 &=& g_+ \bm\alpha - g_- \bm\alpha', \\
\bm\upsilon_2 &=& g_+ \bm\beta - g_- \bm\beta'.\nonumber
\end{eqnarray}

Comparing the two representations (\ref{MPQ}) and (\ref{AB3})
of $M\in Sp(4, R)/U(2)$
we obtain the following expressions for the potentials in terms
of the components ${\upsilon_a}^b$ of the vectors $\bm\upsilon_a$:
\begin{eqnarray} \label{nonde}
f^{-1} &=& g_0 + {\upsilon_0}^0 + {\upsilon_1}^2 + {\upsilon_2}^1,\nonumber\\
\chi f^{-1} &=& {\upsilon_1}^1 - {\upsilon_2}^2 ,\nonumber\\
-v f^{-1} &=& {\upsilon_0}^1 + {\upsilon_2}^0 ,\nonumber\\
u f^{-1} &=& {\upsilon_0}^2 + {\upsilon_1}^0 , \\
{\rm e}^{2\phi}-v^2 f^{-1} &=& g_0 - {\upsilon_0}^0 + {\upsilon_1}^2
- {\upsilon_2}^1,\nonumber\\
\kappa{\rm e}^{2\phi}-uv f^{-1} &=& {\upsilon_1}^1 + {\upsilon_2}^2.\nonumber
\end{eqnarray}
Substituting Eqs. (6.6) and (6.8) one finds for the metric function
$f$ the following expression
\begin{equation}
f^{-1}=\cos\sigma \left(\cosh \sigma + \mu_+\sinh\sigma \right)
+\sin\sigma \left(\mu_0\sinh \sigma + \mu_-\cosh\sigma \right),
\end{equation}
where
\begin{equation}
\mu_0=\Lambda^{-2}\left(|m|^2-|d|^2\right),\quad
\mu_{\pm} = \Lambda^{-1}M \pm \Lambda^{-3}{\rm Re}\left[{\bar q}
(q^2+2md)\right],
\end{equation}
with the normalization factor $\Lambda = |\bm\alpha\wedge\bm\beta|^{1/2}$ for
the $\bm\alpha$, $\bm\beta$ of (\ref{albe}).
We then obtain for the electric and magnetic potentials,
\begin{equation} \label{ndvu}
v+iu=\sqrt{2}f\left\{ qg_+ +(q{\bar m}+{\bar q}d)\Lambda^{-1}g_1 +
{\bar q}(q^2+2md)\Lambda^{-2}g_-\right\}.
\end{equation}
The corresponding dilaton factor is
\begin{equation}
{\rm e}^{2\phi}=v^2f^{-1}+
\cos\sigma \left(\cosh \sigma + \nu_+\sinh\sigma \right)
-\sin\sigma \left(\mu_0\sinh \sigma - \nu_-\cosh\sigma \right),
\end{equation}
where
\begin{equation}
\nu_{\pm}=D\pm\Lambda^{-2}{\rm Re}\left[{\bar m}(q^2+2md)\right].
\end{equation}

Points where $f^{-1}=0$, {\em i.e.}
\begin{equation} \label{sing}
\tan\sigma=-\frac{\cosh\sigma+\mu_+\sinh\sigma}
{\mu_-\cosh\sigma+\mu_0\sinh\sigma}
\end{equation}
correspond to spacetime singularities\footnote{In the case of the
corresponding KK solutions of type 1 \cite{gc}, the function $f^{-1}$
similarly vanishes for a denumerable set of values of $\sigma$. These
four-dimensional spacetime singularities are due to a breakdown of the
Kaluza-Klein dimensional reduction, the five-dimensional metric being
everywhere regular.}. Since $f^{-1}$ enters into the
spacetime metric (\ref{dimred}) as a conformal
factor of the three--space, these singularities (which are also
singularities of the electric and magnetic potentials (\ref{ndvu})) are
point singularities. For a one-center solution $\sigma = k/r$, the
range of $r$ is therefore $]r_1, +\infty[$, where $r_1 > 0$ is the largest
value of $r$ solving eq. (\ref{sing}). The behaviour of the static ($\chi
= 0$) metric near $r = r_1$ is
\begin{equation}
ds^2 \sim \,c^{-2}\rho^{-2/3}\,dt^2 - d\rho^2 - r_1^2c^2\rho^{2/3}\,d\Omega^2
\end{equation}
with $\rho^{2/3} \propto (r - r_1)$. One can show that
all non-radial geodesics, as well as
radial timelike geodesics, are deflected away from $\rho = 0$ ($r = r_1$).
Only radial, lightlike geodesics terminate at the singularity $\rho = 0$,
so this is a very weak singularity. Again, the corresponding
string metric
$d\bar{s}^2 = {\rm e}^{2\phi} ds^2$ turns out to be regular. Generically
the dilaton factor ${\rm e}^{2\phi}$ has the same poles as $f$ so that,
near $r = r_1$,
\begin{equation}
d\bar{s}^2 \sim \frac{a^2}{(r-r_1)^2}\,dt^2 - dr^2 - r_1^2\,d\Omega^2.
\end{equation}
All timelike or lightlike geodesics are deflected away from the throat $r
= r_1$, except for radial lightlike geodesics which
reach the throat after an infinite ``affine time''.

The same local solution (\ref{nonde})
may also be extended to the ranges $r \in ]r_{k+1}, r_k[$, where $\sigma_k
\equiv \sigma(r_k)$ and $\sigma_{k+1}$ are two consecutive roots of
(\ref{sing}); the corresponding spatial sections are compact, with two
point singularities $r = r_k$ and $r = r_{k+1}$. The more general
$n$-center harmonic potential $\sigma = \sum_{j = 1}^n k_j/|\bm{r} -
\bm{r_j}|$ leads to an asymptotically flat/Taub-NUT solution with $p$
point singularities, where $p$ is the number of connected components of
the ``surface'' $\sigma(\bm{r}) = \sigma_1$, as well as compact solutions
with point singularities for $\sigma_k < \sigma(\bm{r}) < \sigma_{k+1}$.
For the corresponding string metrics, the point singularities are replaced
by throats at infinite geodesic distance.

These non-degenerate solutions
may have scalar charges independent of the electromagnetic ones
(up to the BPS condition). Let us consider some particular cases
starting with type 2b ($\bm\alpha,\, \bm\beta$ non--null)
which leads to surprisingly simple formulas.
Here one discovers counterparts of the type 1 extreme Reissner--Nordstr\"om
black holes in the dilaton--axion theory which do not possess
dilaton charges.

1. Electrically charged ``Reissner--Nordstr\"om'':
\begin{equation}
f={\rm e}^{2(\phi-\sigma)}=\frac{{\rm e}^{-\sigma}}
{\sqrt{2}\cos(\sigma-\pi/4)},\quad
v=\frac{\sin\sigma}{\cos(\sigma-\pi/4)},\quad
\sigma=M/r.
\end{equation}
It has $Q=M$, all the other charges being zero. The point  singularity of
the Einstein metric is located at $r_1=4M/3\pi$,
which is also a singularity of the electric and dilaton fields.
Note that the non--zero dilaton field is short range.
The string metric is, as discussed above, regular with a throat $r =
4M/3\pi$ at infinite geodesic distance. In contrast, for the type 1
extreme Reissner--Nordstr\"om solution with $-D = M = Q/\sqrt{2}$, both
the Einstein and string metrics are singular (the coefficient $\gamma$ in
(\ref{Gamma}) vanishes).

2. Magnetically charged ``Reissner--Nordstr\"om'':
\begin{equation}
f={\rm e}^{-2(\phi+\sigma)}=\frac{{\rm e}^{-\sigma}}
{\sqrt{2}\cos(\sigma-\pi/4)},\quad
u=\frac{\sin\sigma}{\cos(\sigma-\pi/4)},\quad
\sigma=M/r,
\end{equation}
which corresponds to $P=M$ and has a similar singularity structure.
However in this case the product $f{\rm e}^{2\phi} = {\rm e}^{-2\sigma}$ is
non--singular, which results in the string metric $d\bar{s}^2$ being
singular (radial geodesics terminate at the point singularity $r = 4M/3\pi$).

These solutions may be subjected to $SO(1,2)\times SO(2)$
transformations to get non--zero values of other charges. In particular,
one can obtain massless asymptotically flat solutions (without NUT's).

3. Massless electrically charged ``geon'' possessing a dilaton charge:
\begin{equation}
f={\rm e}^{2(\phi-\sigma)}=\frac{{\rm e}^{\sigma}}
{\sqrt{2}\cos(\sigma-\pi/4)},\quad
v=\frac{\sin\sigma}{\cos(\sigma-\pi/4)},\quad
\sigma=Q/r.
\end{equation}
It has $D=-Q,\; M=N=A=P=0$. It is worth noting that the gravitational
field is non--zero, but decays as $r^{-2}$.

All the above solutions are static.

4. Massless magnetically charged ``geon'' with axion charge
$A=P$:
\begin{eqnarray}
&&f^{-1}=\cos\sigma \cosh\sigma-\sin\sigma \sinh\sigma,\quad
\chi =f(\sin\sigma \cosh\sigma-\cos\sigma \sinh\sigma),\nonumber\\
&&v=u\tanh\sigma=\sqrt{2}f\sin\sigma \sinh\sigma,\quad
 {\rm e}^{2\phi} =f(\cos^2\sigma+ \sinh^2\sigma)  \\
&&\kappa=f {\rm e}^{-2\phi}(\sin\sigma \cosh\sigma+\cos\sigma\sinh\sigma),
\nonumber
\end{eqnarray}
where $\sigma=P/r$.  This solution possesses short--range electric
and dilaton fields. The singularity corresponds to
\begin{equation}
\cot\sigma=\tanh\sigma.
\end{equation}

Note that for solutions of type 2b the only restriction
on the charges is the BPS condition (\ref{BPS}).
Now consider type 2a solutions (both $\bm\alpha,\, \bm\beta$ are null). In
this case there is a second constraint on the charges, so that the dilaton and
axion charges are functions of the other charges (this dependence
is given by the relations (4.22), (4.23) different from those for
dilaton--axion extreme black holes, (4.12)).

5. Massive symmetric Reissner--Nordstr\"om
 $Q=-P=M/\sqrt{2}$ without dilaton and axion charges:
\begin{eqnarray}
&&f=\frac{{\rm e}^{-\sigma}}
{\sqrt{2}\cos(\sigma-\pi/4)},\quad
u=-v=\frac{\sin\sigma}{\cos(\sigma-\pi/4)},\nonumber\\
&&{\rm e}^{2\phi}=\frac{\cosh\sigma+{\rm e}^{-\sigma}\sin\sigma\cos\sigma}
{\sqrt{2}\cos(\sigma-\pi/4)},\quad
\kappa= \frac{{\rm e}^{-\sigma}\sin\sigma\cos\sigma-\sinh\sigma}
{\cosh\sigma+{\rm e}^{-\sigma}\sin\sigma\cos\sigma},\quad
\sigma=M/r.
\end{eqnarray}
Note that, as in the non--dyon cases, although the dilaton and axion
charges are zero, there still exist short--range dilaton and axion fields.

6. Massless symmetric dyon $P=-Q=A/\sqrt{2}$ without dilaton charge:
\begin{eqnarray}
&&f=\frac{{\rm e}^{\sigma}}
{\sqrt{2}\cos(\sigma-\pi/4)},\quad
u=-v=\frac{\sin\sigma}{\cos(\sigma-\pi/4)},\nonumber\\
&&{\rm e}^{2\phi}=\frac{\cosh\sigma+{\rm e}^\sigma\sin\sigma\cos\sigma}
{\sqrt{2}\cos(\sigma-\pi/4)},\quad
\kappa= \frac{\sinh\sigma+{\rm e}^\sigma\sin\sigma\cos\sigma}
{\cosh\sigma+{\rm e}^\sigma\sin\sigma\cos\sigma},\quad
\sigma=A/r.
\end{eqnarray}
This solution corresponds to an interchange
$\bm\alpha\leftrightarrow\bm\beta$ in the previous solution.

\setcounter{equation}{0}
\section{Discussion}
We have presented a constructive approach to the generation of BPS solutions
to gravity--coupled field theories admitting a $\sigma$--model
representation in three dimensions. This purely bosonic approach has
the advantage of reducing all calculations to algebraic manipulations
with matrices representing relevant cosets (no differential
equations have to be solved). It yields solutions
in a closed form readily parametrized  by physical charges and automatically
symmetric under the dualities preserving asymptotic flatness.
In addition, it provides an algebraic classification
of solutions according to the possible algebraic types of
generating matrices.

For the particular case of dilaton-axion gravity with one vector field (which
was extensively explored during recent years) we have listed
{\em all } stationary asymptotically flat BPS solutions.
These fall into two classes. The first class includes the well--known
multi--extreme black holes and their rather obvious generalizations.
The corresponding Einstein metrics have singular event horizons
(null point singularities). In the case of the
solutions of the new, second class, dilaton and axion charges are not tightly
bound to electric/magnetic charges. Their Einstein metrics
have mild timelike naked singularities. However the
associated string metrics are generically regular for both classes of
solutions.

Both of these classes contain massless states.
In the degenerate case (first class),
the vanishing of mass can be achieved at the expense of the
introduction of NUT's (although for multicenter configurations
the total NUT charge may still be zero).
In the non--degenerate case (new class), there are
strictly asymptoticaly flat massless solutions.

After this work was completed, we received a preprint by Kechkin and
Yurova (hep--th/9604071) in which the results of \cite{g} and \cite{gc}
are also used
to construct classes of stationary BPS solutions to dilaton-axion gravity.
However these authors incorrectly identify the coset to which the
matrices ${\rm e}^{B\sigma}$ belong as $Sp(4,R)/GL(2,R)$. As we have seen,
the correct 
identification
is $Sp(4,R)/(SO(1,2) \times SO(2))$, the SO(2)
transformations corresponding to electric-magnetic duality rotations.
Neither the BPS solutions depending on two harmonic potentials nor our
type 2 solutions are considered in this paper.
\bigskip

{\bf Acknowledgements}
D.G. wishes to thank the Laboratoire de Gravitation et Cosmologie
Relativistes (UPMC/CNRS) for hospitality during his visit while this
work was initiated. He thanks R. Kerner for useful discussions,
and R. Kallosh for stimulating correspondence. He is also grateful
to FAPESP for financial support and to
P. Letelier for his kind invitation to visit Campinas University
while the final version of this paper was completed. This work
was supported in part by the RFBR grant 96--02--18899.

\bigskip
\bigskip
\bigskip

\noindent{\Large \bf APPENDIX A: $Sp(4, R)$ and 1+2 Clifford algebras}
\renewcommand{\theequation}{A.\arabic{equation}}
\setcounter{equation}{0}
\bigskip

The Lie algebra of $Sp(4, R)$ consists of ten real $4\times 4$
matrices $Y$ satisfying the condition
\begin{equation}
Y^TJ+JY=0,
\end{equation}
where $J$ is given by (\ref{J}). For further analysis it is convenient to
introduce the full set of sixteen independent
real $4\times 4$ matrices
as tensor products of two sets of $2\times 2$ matrices
\[
e_{\mu\nu}=\rho_\mu \otimes \sigma_\nu \,, \qquad (\mu, \nu=0, 1, 2, 3) \,,
\]
with
\begin{equation}
\rho_0=\sigma_0=I,\; \rho_1=\sigma_1=\sigma_x,\;
\rho_2=\sigma_2=i\sigma_y,\;  \rho_3=\sigma_3=\sigma_z,\;
\end{equation}
where $\sigma_x,\, \sigma_y,\, \sigma_z$ are the standard Pauli
matrices, and it is assumed that
$\rho_\mu$ act on $2\times 2$ blocks, while $\sigma_\mu$ act inside
these blocks. The following multiplication rules hold
\[
\sigma_2 \sigma_1 =\sigma_3,\quad
\sigma_3 \sigma_2 =\sigma_1,\quad
\sigma_3 \sigma_1 =\sigma_2,\quad
\sigma_1^2=- \sigma_2^2 =\sigma_3^2=I.
\]
In this notation, the matrices introduced in Sect. 3 read $A=e_{03}$,
$K=e_{23}$, while the full Lie algebra  $sp(4, R)$ satisfying (A.1) is
\begin{equation}
sp(4, R)=\left\{e_{02},\, e_{10},\,e_{11},\, e_{13},\, e_{20},\,
e_{21},\, e_{23},\, e_{30},\, e_{31},\, e_{33}\right\}.
\end{equation}

We look for a parametrization of the coset $Sp(4, R)/U(2)$
(choosing as representatives the symmetric symplectic matrices $M$)
of the type
\begin{equation}
M=A{\rm e}^{B\sigma}, \quad A=e_{03}, \quad A^2=I,
\end{equation}
or,
\[ {\rm e}^{B\sigma} = AM. \]
Since $A\in Sp(4, R)/U(2),$ the matrix $B$ belongs to $sp(4, R)$,
\begin{equation} \label{jbj}
JBJ = B^T,
\end{equation}
and is subject to the symmetry condition
\begin{equation} \label{aba}
ABA=B^T.
\end{equation}
The exponential ${\rm e}^{B\sigma}$ is therefore a non--symmetric matrix
which can be shown
to belong to the coset space $Sp(4, R)/(SO(2)\times SO(1, 2))$.
Indeed, solving the equations (\ref{aba}) and (\ref{jbj}) one finds
that $B\in {\cal B}$, with
\begin{equation}
{\cal B}=\left\{e_{02},\, e_{10},\, e_{13},\, e_{21},\, e_{30},\,
e_{33}\right\}.
\end{equation}
The remaining generators of $Sp(4, R)$ form a subalgebra
\begin{equation}
{\cal H'}=\left\{e_{11},\, e_{20},\, e_{23},\, e_{31}\right\}
\end{equation}
which can be identified with $so(2) \oplus so(1, 2)$, where the
$so(2)$ component is $K = e_{23}$, while the $so(1, 2)$ consists of
\begin{equation}
\Sigma_a=\left\{e_{20},\, -e_{11},\, e_{31} \right\}
\end{equation}
($a=0, 1, 2$ is an $so(1, 2)$ index) with $\Sigma_0 \equiv J$, so that
\begin{equation}
\left[ \Sigma_a,\;\Sigma_b \right]=2{\varepsilon_{ab}}^c\,\Sigma_c\,,\qquad
\left[K,\; \Sigma_a \right]=0
\end{equation}
(the three--dimensional Levi--Civita symbol
is defined by $\varepsilon_{012}=1$, and the indices are raised and lowered
with the 1+2 metric $\eta_{ab}=\eta^{ab}={\rm diag}(1, -1, -1)$).
All the elements of ${\cal H'}$ satisfy
\begin{equation} \label{AH}
A{\cal H'}A=-{\cal H'}^T,
\end{equation}
which ensures that the  similarity transformations
of $B$ generated by ${\cal H'}$ preserve the symmetry condition (\ref{aba}).

Inside ${\cal B}$ one can identify two sets of 1+2 Clifford algebras:
\begin{eqnarray}
&& {\Gamma^1}_a= \left\{e_{21},\, -e_{10},\, e_{30}\right\} ,\nonumber\\
&& {\Gamma^2}_a=\left\{e_{02},\, e_{33},\, e_{13}\right\},
\end{eqnarray}
satisfying the anticommutation relations
\begin{equation}
\left\{{\Gamma^1}_a, \;{\Gamma^1}_b\right\}=
\left\{{\Gamma^2}_a, \;{\Gamma^2}_b\right\}=-2\eta_{ab}\, I.
\end{equation}
Their relation to the $so(2)$ subalgebra is clear from
\begin{eqnarray} \label{so2}
&& \left[K,\; {\Gamma^1}_a \right]=-2{\Gamma^2}_a,\nonumber\\
&& \left[K,\; {\Gamma^2}_a \right]=2{\Gamma^1}_a, \\
&& \left[{\Gamma^1}_a,\; {\Gamma^2}_b \right]=2K\eta_{ab}, \nonumber
\end{eqnarray}
that is, $K$ mixes the two sets. Contrary to this, the $so(1, 2)$
does not mix them:
\begin{eqnarray} \label{so12}
&&\left[\Sigma_a,\; {\Gamma^1}_b \right]=
2{\varepsilon_{ab}}^c{\Gamma^1}_c,\nonumber\\
&&\left[\Sigma_a,\; {\Gamma^2}_b \right]=
2{\varepsilon_{ab}}^c{\Gamma^2}_c,\\
&&\left[{\Gamma^1}_a, \;{\Gamma^1}_b\right]=
\left[{\Gamma^2}_a, \;{\Gamma^2}_b\right]=
2{\varepsilon_{ab}}^c \Sigma_c.  \nonumber
\end{eqnarray}

Remarkably, the anticommutator of $\bm\Gamma^1$ and $\bm\Gamma^2$ generates
a third 1+2 Clifford algebra, $\bm\Gamma^0$:
\begin{equation}
\left\{{\Gamma^1}_a, \;{\Gamma^2}_b\right\}=-2{\varepsilon_{ab}}^c\;
{\Gamma^0}_c,\quad
 {\Gamma^0}_a= \left\{e_{03},\, e_{32},\, e_{12}\right\} ,
\end{equation}
consisting of matrices which do not belong to $sp(4, R)$, and satisfying
\begin{equation}
\left\{{\Gamma^0}_a, \;{\Gamma^0}_b\right\}=2\eta_{ab}\, I.
\end{equation}
This new set commutes with the $so(2) \in {\cal H}'$
\begin{equation}
\left[K,\; {\Gamma^0}_a \right]=0,
\end{equation}
while for commutators with the $so(1, 2) \in {\cal H}'$ one has,
similarly to (A.13),
\begin{eqnarray}
&&\left[\Sigma_a,\; {\Gamma^0}_b \right]=
2{\varepsilon_{ab}}^c{\Gamma^0}_c,\nonumber\\
&&\left[{\Gamma^0}_a, \;{\Gamma^0}_b\right]=
-2{\varepsilon_{ab}}^c \Sigma_c.
\end{eqnarray}
Comparing Eqs. (A.11), (A.13), (A.14), (A.17) one can see
that the two--index $SO(1, 2)$ covariant relations
(\ref{anti}) hold indeed. Now, taking the commutator
\begin{equation}
\left[{\Gamma^c}_a,\; {\Gamma^d}_b \right]=
-2{\eta^{cd}}{\varepsilon_{ab}}^e {\Sigma}_e
-2{\varepsilon^{cd}}_e {\eta_{ab}}K^e,
\end{equation}
one finds that the previously introduced $K$ is just the ``time''
component $K=K^0$ of the ``vector''
\begin{equation}
K^a=\{e_{23},\, -e_{01},\, e_{22}\}
\end{equation}
which generates the contravariant $SO(1,2)$  algebra dual to the
``covariant'' $SO(1,2)$  algebra generated by the $\Sigma_a$:
\begin{eqnarray}
&&\left[K^a,\;K^b \right]=
2{\varepsilon^{ab}}_cK^c,\nonumber\\
&&\left[K^a,\;{\Gamma^b}_c \right]=
2{\varepsilon^{ab}}_e{\Gamma^e}_c, \\
&&\left[K^a,\;\Sigma_b \right]=0. \nonumber
\end{eqnarray}
The previously introduced tensor set ${\Gamma^a}_b$ can be expressed as
the tensor product of the two vector sets
\begin{equation}
K^a\Sigma_b={\Gamma^a}_b.
\end{equation}
Together with the unit matrix, the set $\{K^a,\,\Sigma_b,\, {\Gamma^a}_b\}$
constitutes a complete basis of the space of real $4\times 4$ matrices.

\bigskip
\bigskip
\bigskip

\renewcommand{\theequation}{B.\arabic{equation}}
\noindent {\Large \bf APPENDIX B: Sigma model construction of the
Israel-Wilson-Perj\`es solutions}
\setcounter{equation}{0}

\bigskip
We show here how the well-known Israel-Wilson-Perj\`es \cite{IWP}
solutions of the Einstein-Maxwell field equations may be recovered from
the general sigma model construction outlined in Sect. 2.

We first introduce the familiar Ernst potentials \cite{er}
\begin{equation}
{\cal E} = f + i\chi - \overline{\Phi}\Phi \,, \qquad \Phi = \frac{1}{\sqrt{2}}(v
+ iu) \,,
\end{equation}
in terms of which the target space metric (\ref{dl2}) reduces, for vanishing
dilaton and axion fields, $\phi = \kappa = 0$, to
\begin{equation}
dl^{\,2} = \frac{1}{2f^2}|\,d{\cal E} + 2 \overline{\Phi}\,d\Phi|^2 -
\frac{2}{f}\,d\Phi\,d\overline{\Phi} \,.
\end{equation}
This may be identified as the metric of the symmetric space SU(2,1)/S(U(2)
$\times$ U(1)). A matrix representative of this coset may be chosen as
\cite{mg}
\begin{equation} \label{MG}
M = f^{-1} \pmatrix{
1 & \sqrt{2}\,\Phi & i(\overline{\cal E} - {\cal E} + 2\Phi\overline{\Phi})/2 \cr
\sqrt{2}\,\overline{\Phi} & -({\cal E} + \overline{\cal E} - 2\Phi\overline{\Phi})/2 &
-i\sqrt{2}\, {\cal E}\overline{\Phi} \cr
i(\overline{\cal E} - {\cal E} - 2\Phi\overline{\Phi})/2 & i\sqrt{2}\, \overline{\cal E}\Phi &
{\cal E}\overline{\cal E} }.
\end{equation}
This matrix is hermitean, $M^+ = M$, and belongs to SU(2,1), $M^+JM = J$,
${\rm det}M = 1$, with
\begin{equation}
J = \pmatrix{
0 & 0 & -i \cr
0 & 1 & 0 \cr
i & 0 & 0 }.
\end{equation}

An asymptotically flat (or Taub-NUT) solution of the EM field
equations depending on one harmonic potential $\sigma$ may be written in
the exponential form (\ref{AB}), where the constant matrix
\begin{equation}
A = \pmatrix{
1 & 0 & 0 \cr
0 & -1 & 0 \cr
0 & 0 & 1 }
\end{equation}
commutes with $J$. The matrix $B$ belongs to su(2,1),
\begin{equation} \label{con1}
B^+J + JB = 0\,, \qquad {\rm Tr}B = 0\,,
\end{equation}
and is further constrained by the symmetry condition
\begin{equation} \label{con2}
B^+A - AB = 0 \,.
\end{equation}
These conditions also imply ${\rm det}\,B = 0$. The characteristic identity
(\ref{det}) for the matrix $B$ then reduces to
\begin{equation}
B^3 = \frac{1}{2} {\rm Tr}(B^2)\,B \,,
\end{equation}
so that the exponential representation (\ref{AB}) may be rewritten as
\begin{equation}
M = A\,[1 + t^{-1}B\,\sinh t\sigma + t^{-2}B^2\,(\cosh t\sigma - 1)]\,,
\end{equation}
where $t^2 \equiv (1/2)\,{\rm Tr}(B^2)$.

Solutions corresponding to null geodesics in target space satisfy the
further condition
\begin{equation} \label{con3}
{\rm Tr}(B^2) = 0 \,.
\end{equation}
The matrices B solving conditions (\ref{con1}), (\ref{con2}) and
(\ref{con3}) belong to a single matrix type, with $B^3 = 0$, $B^2 \neq 0$,
depending (up to an overall multiplicative constant) on two real
parameters $\alpha$, $\beta$:
\begin{equation}
B = \pmatrix{
2\cos\alpha & \sqrt{2}\,{\rm e}^{-i\beta} & -2\sin\alpha \cr
-\sqrt{2}\,{\rm e}^{i\beta} & 0 & i\sqrt{2}\,{\rm e}^{i\beta} \cr
-2\sin\alpha & i\sqrt{2}\,{\rm e}^{-i\beta} & -2\cos\alpha }.
\end{equation}
Two such matrices $B$ and $B'$ are orthogonal if
\begin{equation} \label{ortho}
{\rm Tr}(BB') \equiv 8[\cos(\alpha - \alpha') - \cos(\beta - \beta')] = 0 \,.
\end{equation}
The commutator $[B, B']$ does not vanish (unless $B' = \pm B$), but
commutes with $B$ and $B'$ {\em iff}
\begin{equation} \label{class}
\alpha' + \beta' = \alpha + \beta \,,
\end{equation}
which also solves (\ref{ortho}). So the matrices $B_a$ belonging to the
class (\ref{class}) may be used to construct a solution (\ref{mupt}) depending
on several harmonic functions $\sigma_a$. However only two of these
matrices are linearly independent. One can choose a basis consisting of
$B$ and of the matrix $C$ defined by $\alpha' = \alpha - \pi/2$, $\beta' =
\beta + \pi/2$, {\em e.g.}
\begin{equation}
C = \pmatrix{
2\sin\alpha & -i\sqrt{2}\,{\rm e}^{-i\beta} & 2\cos\alpha \cr
-i\sqrt{2}\,{\rm e}^{i\beta} & 0 & -\sqrt{2}\,{\rm e}^{i\beta} \cr
2\cos\alpha & \sqrt{2}\,{\rm e}^{-i\beta} & -2\sin\alpha }.
\end{equation}
This choice, such that
\begin{equation}
BC + CB = 0 \,, \quad C^2 = B^2 \,,
\end{equation}
leads to the simple form for the solution depending on two potentials
\begin{equation} \label{solEM}
M = A\,{\rm e}^{B\sigma + C\tau} = A[1 + B\sigma + C\tau +
\frac{1}{2}B^2(\sigma^2 + \tau^2)]\,.
\end{equation}

Combining the two real potentials $\sigma$, $\tau$ into a single
complex harmonic potential
\begin{equation}
\zeta = \sigma - i\tau \,,
\end{equation}
we recover from (\ref{solEM}) and (\ref{MG}) the Ernst potentials
\begin{equation} \label{EPhi}
{\cal E} = \frac{1 - {\rm e}^{i\alpha}\zeta}{1 + {\rm e}^{i\alpha}\zeta}
\,, \quad \Phi = \frac{{\rm e}^{-i\beta}\zeta}{1 + {\rm e}^{i\alpha}\zeta}
\,,
\end{equation}
leading to the gravitational potentials
\begin{eqnarray}
f^{-1} & = & |1 + {\rm e}^{i\alpha}\zeta|^2 = 1 + 2(\sigma\,\cos\alpha +
\tau\,\sin\alpha) + \sigma^2 + \tau^2 \,, \\
\chi & = & 2f\,(\tau\,\cos\alpha - \sigma\,\sin\alpha) \,,
\end{eqnarray}
to be compared with the solution (\ref{sol1}) for dilaton-axion gravity.
Eq. (\ref{EPhi})
implies the linear relation between the Ernst potentials
\begin{equation}
\Phi = \frac{1}{2}\,{\rm e}^{-i(\alpha + \beta)} (1 - {\cal E}) \,,
\end{equation}
which is equation (24) of Israel and Wilson \cite{IWP}, while their
harmonic potential $2/(1 + {\cal E}) - 1$ is our ${\rm e}^{i\alpha}\zeta$.

\end{document}